\documentclass[prd,twocolumn,groupedaddress,showpacs,nofootinbib]{revtex4}
\usepackage{graphicx}
\usepackage{dcolumn}
\usepackage{amssymb}
\usepackage{mathrsfs}
\usepackage{amsmath}
\usepackage{epsfig}
\usepackage[dvips]{color}
\usepackage{hhline}

\begin{document}

\title{Neutrinoless double beta decay in the left-right symmetric models for linear seesaw}

\author{Pei-Hong Gu}

\email{peihong.gu@sjtu.edu.cn}

\affiliation{Department of Physics and Astronomy, Shanghai Jiao Tong University, 800 Dongchuan Road, Shanghai 200240, China}

\begin{abstract}

In a class of left-right symmetric models for linear seesaw, a neutrinoless double beta decay induced by the left- and right-handed charged currents together will only depend on the breaking details of left-right and electroweak symmetries. This neutrinoless double beta decay can reach the experimental sensitivities if the right-handed charged gauge boson is below the 100 TeV scale.

\end{abstract}

\pacs{14.60.Pq, 14.60.St, 12.60.Cn, 12.60.Fr}

\maketitle

\section{Introduction}

The phenomena of neutrino oscillations have been established by atmospheric and solar neutrino measurements, and also confirmed by accelerator and reactor neutrino experiments \cite{olive2014}. This fact implies the need for massive and mixing neutrinos and hence new physics beyond the $SU(3)_c^{}\times SU(2)_L^{}\times U(1)^{}_{Y}$ standard model (SM). Furthermore, the cosmological observations have indicated that the neutrino masses should be in a sub-eV range \cite{olive2014}. The seesaw \cite{minkowski1977,yanagida1979,grs1979,ms1980} mechanism provides a natural way to the tiny neutrino masses \cite{minkowski1977,yanagida1979,grs1979,ms1980,mw1980,sv1980,cl1980,lsw1981,ms1981,flhj1989,ma1998}. For realizing a seesaw scenario, one can simply extend the SM by introducing some heavy particles with lepton number violation of two units. Another appealing scheme is to consider the $SU(3)_c^{}\times SU(2)_L^{}\times SU(2)_R^{}\times U(1)^{}_{B-L}$ left-right symmetric models (LRSMs) \cite{ps1974,mp1975,mp1975-2,ms1975}, where the lepton number violation and the heavy particles can naturally appear after the left-right symmetry is spontaneously broken down to the electroweak symmetry. For example, the LRSMs with $[SU(2)]$-triplet and bidoublet Higgs scalars can accommodate a type-I+II seesaw \cite{ms1980,ms1981}.

It should be noted the neutrinos are neutral and hence are allowed to have a Majorana nature \cite{majorana1937}. One thus can expect a neutrinoless double beta decay ($0\nu\beta\beta$) \cite{furry1939} process mediated by the Majorana electron neutrinos. This $0\nu\beta\beta$ process is determined by the Majorana neutrino mass matrix \cite{rodejohann2011,bgmr2013}. Alternatively, the $0\nu\beta\beta$ process can come from other lepton number violating interactions. For example, there have been a number of people studying the $0\nu\beta\beta$ processes in the TeV-scale LRSMs for the type-I+II seesaw  \cite{ms1980,ms1981,tnnsv2010,pp2012,ddkv2012,app2013,br2013,hl2014,dgpss2015,bgm2015,glp2015,abm2015,bbgm2015}. Specifically, people need do some assumptions on the scale of the right-handed neutrinos as well as the mixing between the left- and right-handed neutrinos to quantitatively analyze the $0\nu\beta\beta$ processes involving the right-handed charged currents. One may consider other models which induce a testable $0\nu\beta\beta$ process at tree level and then give a negligible contribution to the neutrino masses at loop level \cite{sv1982}. The $0\nu\beta\beta$ processes thus can be free of the neutrino masses \cite{bm2002,gu2011,bhow2013}. But these models contain quite a few arbitrary parameters.

In this paper we shall study the $0\nu\beta\beta$ processes in a class of LRSMs for the so-called linear seesaw \cite{barr2003,mrv2005}, where the left-handed neutrinos have a Majorana mass matrix proportional to the Dirac mass term between the left- and right-handed neutrinos. Our illustration will show that a $0\nu\beta\beta$ process induced by the left- and right-handed charged currents together is irrelevant to any masses and mixing of the left- and right-handed neutrinos, instead it is only dependent on the vacuum expectation values (VEVs) of some Higgs scalars. This $0\nu\beta\beta$ process can arrive at a testable level if the right-handed charged gauge boson is below the 100 TeV scale.

\section{Linear seesaw}

We start with the original LRSM, where the Higgs multiplets include two $SU(2)$ doublets and one $SU(2)_L^{}\times SU(2)_R^{}$ bidoublet,
\begin{eqnarray}
\label{higgs1}
&&\chi^{}_L(1,2,1,-1),~~\chi^{}_R(1,1,2,-1),~~\Phi(1,2,2,0)~~\textrm{with}\nonumber\\
[2mm]
&&\chi^{}_{L,R}\!=\!\left[\begin{array}{l}\chi^{0}_{L,R}\\
[2mm]
\chi^{-}_{L,R}\end{array}\right]\!,~~\Phi\!=\!\left[\begin{array}{ll}\phi^{0}_{1}&\phi^{+}_{2}\\
[2mm]
\phi^{-}_{1}&\phi^{0}_{2}\end{array}\right]\equiv [\phi_1^{}~~\tilde{\phi}_2^{}].
\end{eqnarray}
Here and thereafter the brackets following the fields describe the transformations under the $SU(3)^{}_{c}\times SU(2)^{}_{L}\times SU(2)_R^{}\times U(1)^{}_{B-L}$ gauge groups. In the fermion sector, we have three generations of $[SU(2)]$-doublet fermions,
\begin{eqnarray}
\label{fermion1}
\!\!\!\!\begin{array}{l}q^{}_{L}(3,2,1,+\frac{1}{3})\end{array}\!=\!\left[\begin{array}{l}u^{}_{L}\\
d^{}_{L}\end{array}\right]\!,\!\!&&\!\!
\begin{array}{l}q^{}_{R}(3,1,2,+\frac{1}{3})\end{array}\!=\!\left[\begin{array}{l}u^{}_{R}\\
[2mm]
d^{}_{R}\end{array}\right]\!,\nonumber\\
[2mm]
\!\!\!\!\begin{array}{l}l^{}_{L}(1,2,1,-1)\end{array}\!=\!\left[\begin{array}{l}\nu^{}_{L}\\
[2mm]
e^{}_{L}\end{array}\right]\!,\!\!&&\!\!
\begin{array}{l}l^{}_{R}(1,1,2,-1)\end{array}
\!=\!\left[\begin{array}{l}\nu^{}_{R}\\
e^{}_{R}\end{array}\right]\!.
\end{eqnarray}

One can analyze the symmetry breaking in details from the fully renormalizable scalar potential which contains
\begin{eqnarray}
V\supset-\rho\chi^\dagger_L\Phi\chi_R^{}
-\tilde{\rho}\chi^\dagger_L\tilde{\Phi}\chi_R^{}+\textrm{H.c.}\,.
\end{eqnarray}
The acceptable breaking pattern is $SU(2)_L^{}\times SU(2)_R^{}\times U(1)_{B-L}^{}\stackrel{\langle\chi_R^{}\rangle}{\longrightarrow} SU(2)_L^{}\times U(1)_{Y}^{}\stackrel{\langle\varphi\rangle}{\longrightarrow} U(1)_{em}^{}$. Here the $\varphi$ scalar is a linear combination of three $SU(2)_L^{}$ doublets $\phi_{1,2}$ and $\chi_L^{}$,
\begin{eqnarray}
\varphi\equiv \frac{\langle\phi_1^{}\rangle\phi_1^{}+\langle\phi_2^{}\rangle\phi_2^{}+\langle\chi_L^{}\rangle\chi_L^{}}
{\sqrt{\langle\phi_1^{}\rangle^{2}_{}+\langle\phi_2^{}\rangle^2_{}+\langle\chi_L^{}\rangle^2_{}}}~\textrm{with}~\langle\varphi\rangle=174\,\textrm{GeV}\,.
\end{eqnarray}
The allowed Yukawa interactions are
\begin{eqnarray}
\label{yukawa1}
\mathcal{L}\!\supset\!-y_q^{}\bar{q}_{L}^{}\Phi q_{R}^{}\!-\!\tilde{y}_q^{}\bar{q}_{L}^{}\tilde{\Phi} q_{R}^{}\!-\!y_l^{}\bar{l}_{L}^{}\Phi l_{R}^{}\!-\!\tilde{y}_l^{}\bar{l}_{L}^{}\tilde{\Phi} l_{R}^{}\!+\!\textrm{H.c.},
\end{eqnarray}
from which one can read the fermion masses,
\begin{eqnarray}
\label{cfmass}
\mathcal{L}\!&\supset&\!-m_u^{}\bar{u}_L^{}u_R^{}-m_d^{}\bar{d}_L^{}d_R^{}-m_e^{}\bar{e}_L^{}e_R^{}-m_D^{}\bar{\nu}_L^{}\nu_R^{}+\textrm{H.c.}\nonumber\\
[2mm]
\!&&\!\textrm{with}~~\left\{\begin{array}{lcl}m_{u(D)}^{}&=&y_{q(l)}^{}\langle\phi_1^{}\rangle+\tilde{y}_{q(l)}^{}\langle\phi_2^{}\rangle\,,\\
[2mm]
m_{d(e)}^{}&=&y_{q(l)}^{}\langle\tilde{\phi}_2^{}\rangle+\tilde{y}_{q(l)}^{}\langle\tilde{\phi}_1^{}\rangle\,.\end{array}\right.
\end{eqnarray}
Clearly one must fine tune some Yukawa couplings to make the Dirac neutrino masses below the eV scale.

It has been shown that the original LRSM can generate the tiny neutrino masses in a natural way if it is extended by three gauge-singlet fermions \cite{alsv1996,alsv1996-2},
\begin{eqnarray}
\label{fermion2}
\xi_{R}^{}(1,1,1,0)\,,
\end{eqnarray}
which have the Yukawa couplings,
\begin{eqnarray}
\label{yukawa2}
\mathcal{L}\supset-f_L^{}\bar{l}_L^{}\chi_L^{}\xi_R^{}-f_R^{}\bar{l}_R^{c}\chi_R^{\ast}\xi_R^{}+\textrm{H.c.}\,.
\end{eqnarray}
In principle, the singlet fermions are also allowed to have a gauge-invariant Majorana mass term. Here we simply assume that this term does not exist, i.e.
\begin{eqnarray}
\mathcal{L}~/\!\!\!\!\!\!\supset-\frac{1}{2}m_\xi^{}\bar{\xi}_R^{c}\xi_R^{}+\textrm{H.c.}~~\textrm{with}~~m_\xi^{}=m_\xi^T\,.
\end{eqnarray}
We will give a realistic model to explain this assumption in Sec. IV. Under this assumption, we can obtain three Dirac pairs composed of the right-handed neutrinos $\nu_R^{}$ and the singlet fermions $\xi_R^{}$ at the electroweak level.

By choosing the discrete left-right symmetry to be the $CP$ under which the fermions and the scalars have the following transformations,
\begin{eqnarray}
\!\!\!\!\!\!\!\!\!\!\!\!&&q_{L}^{}\!\stackrel{CP}{\longleftrightarrow}\!q_{R}^{c},~l_{L}^{}\!\stackrel{CP}{\longleftrightarrow}\!l_{R}^{c},~
\xi_{R}^{}\!\stackrel{CP}{\longleftrightarrow}\!\xi_{R}^{},
~\chi_L^{}\!\stackrel{CP}{\longleftrightarrow}\!\chi_{R}^{\ast},~\Phi\!\stackrel{CP}{\longleftrightarrow}\!\Phi^T_{},\nonumber\\
\!\!\!\!\!\!\!\!\!\!\!\!&&
\end{eqnarray}
one can simplify the Yukawa couplings (\ref{yukawa1}) and (\ref{yukawa2}) as
\begin{eqnarray}
y_q^{}=y_q^T,~\tilde{y}_q^{}=\tilde{y}_q^T,~y_l^{}=y_l^T,~\tilde{y}_l^{}=\tilde{y}_l^T,~f_L^{}=f_R^{}=f.
\end{eqnarray}
The charged fermion mass matrices in Eq. (\ref{cfmass}) thus should be symmetric, i.e. $m_u^{}=m_u^T$, $m_d^{}=m_d^T$ and $m_e^{}=m_e^T$. As for the neutral fermions, their masses are
\begin{eqnarray}
\label{lrsmass}
\mathcal{L}\!\!&\supset&\!\!-\frac{1}{2}[\begin{array}{lll}\bar{\nu}_L^{}&\bar{\nu}_R^{c}&\bar{\xi}_R^{c} \end{array}]\!\!\left[\begin{array}{ccc}0&m_D^{}& f \langle \chi_L^{}\rangle\\
[2mm]
m_D^{T} & 0 & f\langle \chi_R^{}\rangle\\
[2mm]
f_{}^T\langle \chi_L^{}\rangle & f_{}^T\langle \chi_R^{}\rangle & 0 \end{array}\right]\!\!\!\!\left[\begin{array}{lll}\nu_L^{c}\\
[2mm]
\nu_R^{}\\
[2mm]
\xi_R^{} \end{array}\right]\nonumber\\
[2mm]
\!\!&&+\textrm{H.c.}~~\textrm{with}~~m_D^{}=m_D^T\,.
\end{eqnarray}
For $f\langle \chi_R^{}\rangle\gg m_D^{}$ and $ \langle\chi_R^{}\rangle\gg \langle \chi_L^{}\rangle $, the left-handed neutrinos can get a Majorana mass term through the seesaw mechanism, i.e.
\begin{eqnarray}
\mathcal{L}\supset-\frac{1}{2}m_\nu^{}\bar{\nu}_L^{}\nu_L^c+\textrm{H.c.}~~\textrm{with}~~m_\nu^{}=-2m_D^{}\frac{\langle\chi_L^{}\rangle}{\langle\chi_R^{}\rangle}\,.
\end{eqnarray}
Remarkably, the Majorana mass matrix of the left-handed neutrinos is proportional to the Dirac mass term between the left- and right-handed neutrinos. This formula of the neutrino masses is referred to as a linear seesaw. Usually one takes a sizable Dirac mass term $m_D^{}$ and a small factor $\langle\chi_L^{}\rangle/\langle\chi_R^{}\rangle$ to generate the tiny neutrino masses $m_\nu^{}$. In the following we will show a small $m_D^{}$ and a sizable $\langle\chi_L^{}\rangle/\langle\chi_R^{}\rangle$ is interesting to the $0\nu\beta\beta$ process. In Sec. IV, we will also explain the small $m_D^{}$ in a realistic model.

\begin{figure}
\vspace{4.2cm} \epsfig{file=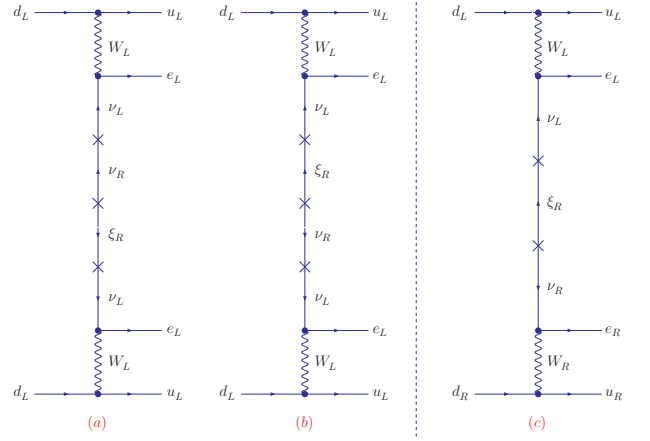, bbllx=5.7cm, bblly=6.0cm,
bburx=15.7cm, bbury=16cm, width=4cm, height=4cm, angle=0,
clip=0} \vspace{-2.3cm} \caption{\label{neutrinoless} The dominant neutrinoless double beta decay processes in the left-right symmetric models for linear seesaw. The diagrams (a) and (b) are only from the left-handed currents, while the diagram (c) is induced by the left- and right-handed currents together. }
\end{figure}

\section{Neutrinoless double beta decay}

As shown in Fig. \ref{neutrinoless}, the left-handed neutrinos $\nu_L^{}$, the right-handed neutrinos $\nu_R^{}$ and the singlet fermions $\xi_R^{}$ with the masses (\ref{lrsmass}) can mediate some $0\nu\beta\beta$ processes in association with the charged currents,
\begin{eqnarray}
\mathcal{L}&\supset&\frac{g}{\sqrt{2}}(\bar{u}_L^{}\gamma^\mu_{}d_L^{}W_{L\mu}^{+}+\bar{e}_L^{}\gamma^\mu_{}\nu_L^{}W_{L\mu}^{-}\nonumber\\
[2mm]
&&+\bar{u}_R^{}\gamma^\mu_{}d_R^{}W_{R\mu}^{+}+\bar{e}_R^{}\gamma^\mu_{}U_R^{}\hat{\nu}_R^{}W_{R\mu}^{-})\,.
\end{eqnarray}
Here we have rotated the right-handed neutrinos and the singlet fermions to diagonalize their masses, i.e.
\begin{eqnarray}
\nu_R^{}=U_R^{}\hat{\nu}_R^{}\,,~~\xi_R^{}=U_\xi^{}\hat{\xi}_R^{}\,,~~\hat{f}=U_R^{T}f U_\xi^{}\,,
\end{eqnarray}
and hence have rewritten the mass matrix in Eq. (\ref{lrsmass}) by
\begin{eqnarray}
\left[\begin{array}{ccc}0&m_D^{}U_R^{}& \hat{f} \langle\chi_L^{}\rangle U_R^\dagger\\
[2mm]
U_R^T m_D^{} & 0 & \hat{f} \langle\chi_R^{}\rangle\\
[2mm]
U_R^\ast\hat{f} \langle\chi_L^{}\rangle  & \hat{f} \langle\chi_R^{}\rangle & 0 \end{array}\right]\,.
\end{eqnarray}
By integrating out the heavy Dirac fermions composed of $\hat{\nu}_R^{}$ and $\hat{\xi}_R^{}$, we can derive the effective $0\nu\beta\beta$ operators,
\begin{eqnarray}
\!\!\!\!\!\!\!\!\mathcal{L}\!\!&\supset&\!\!16\,G_F^2\bar{u}_L^{}\gamma^\mu_{}d_L^{}\bar{e}_L^{}\gamma_\mu^{}\frac{(m_\nu^{})_{ee}^{}}{q^2_{}}\gamma_\nu^{}e_L^c \bar{u}_L^{}\gamma^\nu_{}d_L^{}\nonumber\\
[2mm]
\!\!\!\!\!\!\!\!\!\!&&\!\!+ 8\frac{\langle\chi_L^{}\rangle}{\langle\chi_R^{}\rangle}\frac{M_{W_L^{}}^{2}}{M_{W_R^{}}^{2}}
G_F^2\bar{u}_L^{}\gamma^\mu_{}d_L^{}\bar{e}_L^{}\gamma_\mu^{}\frac{q\!\!\!\diagup}{q^2_{}}\gamma_\nu^{}e_R^c \bar{u}_R^{}\gamma^\nu_{}d_R^{}\,,
\end{eqnarray}
where the first and second terms (with $q$ being the transfer momentum at the lepton vertex) are respectively from the diagrams (a)+(b) and the diagram (c) in Fig. \ref{neutrinoless}. Here we have ignored the mixing between the $W_L^{}$ and $W_R^{}$ gauge bosons for $\langle\chi_R^{}\rangle\gg\langle\varphi\rangle$. In this case the $W_L^{}$ and $W_R^{}$ masses are simply given by
\begin{eqnarray}
M_{W_L^{}}^2=\frac{1}{2}g^2_{}\langle\varphi\rangle^2_{}\,,~~
M_{W_R^{}}^2=\frac{1}{2}g^2_{}\langle\chi_R^{}\rangle^2_{}\,.
\end{eqnarray}

One can take the $W_L^{}\!-\!W_R^{}$ mixing into account for more $0\nu\beta\beta$ processes. In general, a reasonably large $W_L-W_R$ mixing can significantly contribute to the $0\nu\beta\beta$ processes. However, the $W_L^{}\!-\!W_R^{}$ mixing can be as small as 
\begin{eqnarray}
\label{wlwr}
\zeta &\sim& \frac{3g^2_{}}{32\pi^2_{}}\frac{m_t^{}m_b^{}}{M_{W_R}^2}=\mathcal{O}(10^{-6}-10^{-10})\nonumber\\
&&\textrm{for}~~M_{W_R}=\mathcal{O}(1-100\,\textrm{TeV})\,,
\end{eqnarray}
which is a one-loop contribution dominated by the top and bottom quarks. In this case, we can neglect the $0\nu\beta\beta$ processes from the $W_L^{}\!-\!W_R^{}$ mixing. Actually, we will demonstrate a realistic model to explain the absence of the Majorana mass term of the singlet fermions as well as the smallness of the Dirac mass term between the left- and right-handed neutrinos. In this realistic model, the $W_L^{}\!-\!W_R^{}$ mixing is dominated by Eq. (\ref{wlwr}).

The half-life of the $0\nu\beta\beta$ processes is calculated by
\begin{eqnarray}
\label{prediction}
\!\!\!\!\frac{1}{T_{1/2}^{0\nu}}\!=\!G_{01}^{0\nu}\!\!\left[\!|M_{LL}^{0\nu}|^2_{}\frac{|(m_{\nu}^{})_{ee}^{}|^2_{}}{m_e^2}\!
+\!|M_{LR}^{0\nu}|^2_{}\frac{1}{4}\frac{\langle\chi_L^{}\rangle^2_{}}{\langle\chi_R^{}\rangle^2_{}}
\frac{M_{W_L^{}}^4}{M_{W_R^{}}^4}\!\right]\!\!,
\end{eqnarray}
with $G_{01}^{0\nu}$ being the phase space factor and $M_{LL,LR}^{0\nu}$ being the nuclear matrix elements. As the Yukawa couplings $|y_{q(l)}^{}|, |\tilde{y}_{q(l)}^{}|< \sqrt{4\pi}$ are allowed by the perturbation requirement, we can take $\langle\chi_L^{}\rangle$ very close to $\langle\varphi\rangle$, i.e. $\langle\chi_L^{}\rangle\simeq \langle\varphi\rangle$. Accordingly the factor $\langle\chi_L^{}\rangle/\langle\chi_R^{}\rangle$ can approximate to $\langle\chi_L^{}\rangle/\langle\chi_R^{}\rangle=\langle\chi_L^{}\rangle/ \langle\varphi\rangle \cdot M_{W_L^{}}^{}/M_{W_R^{}}^{}\simeq M_{W_L^{}}^{}/M_{W_R^{}}^{}$. For the $_{}^{136}\textrm{Xe}$ isotope with $G_{01}^{0\nu}=3.56\times 10^{-14}_{}\,\textrm{yr}^{-1}_{}$, $M_{LL}^{}=1.57-3.85$ and $M_{LR}^{}=1.92-2.49$ \cite{psvf1996,ki2012}, we then find $T_{1/2}^{0\nu}(_{}^{136}\textrm{Xe})=3.4\times 10^{25}_{}\,\textrm{yr}$ for $M_{W_R^{}}^{}=8.2-8.9\,\textrm{TeV}$ and $(m_\nu^{})_{ee}^{}=0$. As for the $_{}^{76}\textrm{Ge}$ isotope with $G_{01}^{0\nu}=5.77\times 10^{-15}_{}\,\textrm{yr}^{-1}_{}$, $M_{LL}^{}=2.58-6.64$ and $M_{LR}^{}=1.75-3.76$ \cite{psvf1996,ki2012}, the prediction is $T_{1/2}^{0\nu}(_{}^{76}\textrm{Ge})=5.2\times 10^{25}_{}\,\textrm{yr}$ for $M_{W_R^{}}^{}=10.1-12.8\,\textrm{TeV}$ and $(m_\nu^{})_{ee}^{}=0$. Currently the experimental limits are $T_{1/2}^{0\nu}(_{}^{136}\textrm{Xe})>3.4\times 10^{25}_{}\,\textrm{yr}$ from the KamLAND-Zen collaboration \cite{gando2012} and $T_{1/2}^{0\nu}(_{}^{76}\textrm{Ge})>5.2\times 10^{25}_{}\,\textrm{yr}$ from the GERDA collaboration \cite{agostini2016}. The $0\nu\beta\beta$ half-life sensitivity is expected to improve in the future, such as $T_{1/2}^{0\nu}(_{}^{136}\textrm{Xe})>8\times 10^{26}_{}\,\textrm{yr}$ \cite{auger2012} and $T_{1/2}^{0\nu}(_{}^{76}\textrm{Ge})>6\times 10^{27}_{}\,\textrm{yr}$ \cite{ackermann2013,abgrall2013}. The half-life in Eq. (\ref{prediction}) can arrive at a testable level $T_{1/2}^{0\nu}(_{}^{136}\textrm{Xe})<8\times 10^{26}_{}\,\textrm{yr}$ for $M_{W_R^{}}^{}<14.2-15.5\,\textrm{TeV}$ and $T_{1/2}^{0\nu}(_{}^{76}\textrm{Ge})<6\times 10^{27}_{}\,\textrm{yr}$ for $M_{W_R^{}}^{}<15.6-20.1\,\textrm{TeV}$. Note this is true even if the electron neutrino has a zero Majorana mass $(m_\nu^{})_{ee}^{}=0$.

In the above demonstration, the mixing between the left-handed neutrinos and the singlet fermions is of the order of $\langle\chi_L^{}\rangle/\langle\chi_R^{}\rangle\simeq M_{W_L^{}}^{}/M_{W_R^{}}^{}$. Such mixing will affect the invisible decay width of the $Z$ boson if the Dirac pairs composed of the singlet fermions and the right-handed neutrinos are heavier than the $Z$ boson. Compared with its SM value, the $Z$ invisible decay width will be reduced by a factor $1-\langle\chi_L^{}\rangle^2_{}/\langle\chi_R^{}\rangle^2_{}$. So, the present parameter choice can be consistent with the precise measurement $\Gamma_Z^{}(\textrm{invisible})=499.0\pm 1.5\,\textrm{MeV}$ \cite{olive2014} even if the singlet fermions and the right-handed neutrinos are heavier than the $Z$ boson.

\section{A realistic model}

The above demonstrations are based on two assumptions: (i) the absence of the Majorana mass term of the singlet fermions; (ii) the smallness of the Dirac mass term between the left- and right-handed neutrinos. We now present a realistic model to naturally account for these assumptions. Besides the previous fermions (\ref{fermion1}) and (\ref{fermion2}), this model contains other three generations of $[SU(2)]$-singlet fermions,
\begin{eqnarray}
\label{fermion3}
\!\!\!\!\begin{array}{c}U_{L,R}^{}(3,1,1,+\frac{4}{3}),\,D_{L,R}^{}(3,1,1,-\frac{2}{3}),\,E_{L,R}^{}(1,1,1,-2).\end{array}
\end{eqnarray}
The Higgs sector (\ref{higgs1}) is also enlarged to be,
\begin{eqnarray}
&&\sigma(1,1,1,0),~~\Phi(1,2,2,0)\equiv[\phi_1^{}~\tilde{\phi}_2^{}],\nonumber\\
[2mm]
&&\chi_{La}^{}(1,2,1,-1),~~\chi_{Ra}(1,1,2,-1)~~(a=1,2).
\end{eqnarray}
There is a $U(1)_{PQ}^{}$ global symmetry under which the fermions and the scalars carry the charges as shown in Table \ref{pqcharge}. We will show later this $U(1)_{PQ}^{}$ is indeed a Peccei-Quinn (PQ) symmetry \cite{pq1977,weinberg1978,wilczek1978} for solving the strong CP problem.

\begin{table}
\vspace{2mm}
\begin{center}
\begin{tabular}{|c|c|c|}  \hline &&\\[-2.5mm]
$F_{L}^{}\!\stackrel{CP}{\longleftrightarrow}\!F_{R}^{c}$&$(\chi_{L1}^{},\chi_{L2}^{\ast},\Phi,\sigma)\!
\stackrel{CP}{\longleftrightarrow}\!(\chi_{R1}^{\ast},\chi_{R2}^{},\Phi^T_{},\sigma)$&$\xi_R^{}\!\stackrel{CP}{\longleftrightarrow}\!\xi_R^{}$\\
&&\\[-2.5mm]
\hline&&\\[-2.5mm]
$+1$&$+2$&$+3$
\\[1mm]\hline
\end{tabular}
\caption{\label{pqcharge} The $U(1)_{PQ}^{}$ charges. Here we have denoted $F_L^{}=(q_L^{},l_L^{},U_L^{},D_L^{},E_L^{})$ and $F_R^{}=(q_R^{},l_R^{},U_R^{},D_R^{},E_R^{})$.}
\end{center}
\end{table}

Under the discrete CP symmetry and the global PQ symmetry, the scalar potential should include
\begin{eqnarray}
\!\!\!\!\!\!V\!\!\!&\supset&\!\!\! \kappa_{11}^{}\sigma\chi_{L1}^\dagger\Phi\chi_{R1}^{}\!+\!\kappa_{12}^{}\sigma^\ast_{}(\chi_{L1}^\dagger\Phi\chi_{R2}^{}\!+\!\chi_{R1}^T\Phi^T_{}\chi_{L2}^{\ast})
\nonumber\\
[2mm]
\!\!\!\!\!\!\!\!\!&&\!\!\!+\tilde{\kappa}_{22}^{}\sigma^\ast_{}\chi_{L2}^\dagger\tilde{\Phi}\chi_{R2}^{}\!+\tilde{\kappa}_{12}\sigma(\chi_{L1}^\dagger\tilde{\Phi}\chi_{R2}^{}\!+\!\chi_{R1}^T\tilde{\Phi}^T_{}\chi_{L2}^{\ast})
\nonumber\\
[2mm]
\!\!\!\!\!\!\!\!\!&&\!\!\!+\zeta\sigma^2_{}\textrm{Tr}(\Phi^\dagger_{}\tilde{\Phi})\!+\!\lambda_{12}^{}\sigma^2_{} (\chi_{L1}^\dagger\chi_{L2}^{}\!+\!\chi_{R1}^T\chi_{R2}^{\ast})
\!+\!\textrm{H.c.}.
\end{eqnarray}
The gauge-singlet scalar $\sigma$ is responsible for the spontaneous PQ symmetry breaking, i.e.
\begin{eqnarray}
\sigma=\frac{1}{\sqrt{2}}(f_{PQ}^{}+h_{PQ}^{})\textrm{exp}(ia/f_{PQ}^{})\,,
\end{eqnarray}
with $f_{PQ}^{}=\sqrt{2}\langle\sigma\rangle$ being a VEV, $h_{PQ}^{}$ being a Higgs boson and $a$ being a Goldstone boson.
Subsequently, we can define an $[SU(2)_R^{}]$-doublet scalar and an $[SU(2)_L^{}]$-doublet scalar,
\begin{eqnarray}
\!\!\!\!\chi_R^{}\!\equiv\! \frac{\sum_{a}^{}\langle\chi_{Ra}^{}\rangle\chi_{Ra}^{}}
{\sqrt{\sum_{a}^{}\langle\chi_{Ra}^{}\rangle^2_{}}},~~
\varphi\!\equiv\! \frac{\sum_{a}^{}(\langle\phi_a^{}\rangle\phi_a^{}+\langle\chi_{La}^{}\rangle\chi_{La}^{})}
{\sqrt{\sum_{a}^{}(\langle\phi_a^{}\rangle^{2}_{}+\langle\chi_{La}^{}\rangle^2_{})}},
\end{eqnarray}
to respectively drive the left-right and electroweak symmetry breaking. We would like to emphasize that the VEVs of the $[SU(2)_L^{}\times SU(2)_R^{}]$-bidoublet Higgs scalar $\Phi$ are seesaw-suppressed,
\begin{eqnarray}
\langle\phi_1^{}\rangle&=&-\frac{\langle\sigma\rangle}{M_{\phi_1}^2}[\kappa_{11}^{}\langle\chi_{L1}^{}\rangle\langle\chi_{R1}^{}\rangle\nonumber\\
[2mm]
&&+\kappa_{12}^{}(\langle\chi_{L1}^{}\rangle\langle\chi_{R2}^{}\rangle+\langle\chi_{L2}^{}\rangle\langle\chi_{R1}^{}\rangle)]
\ll \langle\varphi\rangle\,,\nonumber\\
[2mm]
\langle\phi_2^{}\rangle&=&-\frac{\langle\sigma\rangle}{M_{\phi_2}^2}[\tilde{\kappa}_{22}^{}\langle\chi_{L2}^{}\rangle\langle\chi_{R2}^{}\rangle\nonumber\\
[2mm]
&&
+\tilde{\kappa}_{12}^{}(\langle\chi_{L1}^{}\rangle\langle\chi_{R2}^{}\rangle+\langle\chi_{L2}^{}\rangle\langle\chi_{R1}^{}\rangle)]\ll \langle\varphi\rangle\,.
\end{eqnarray}
Actually for $\langle\chi_{L1,2}^{}\rangle<\langle\varphi\rangle=174\,\textrm{GeV}$, it is easy to give $\langle\phi_{1,2}^{}\rangle=\mathcal{O}(1-100\,\textrm{eV})$ by inputting $\langle\sigma\rangle=\mathcal{O}(10^{10-12}_{}\,\textrm{GeV})$, $M_{\phi_{1,2}}^{}=\mathcal{O}(10^{12-13}_{}\,\textrm{GeV})$ and $\langle\chi_{R1,2}^{}\rangle=\mathcal{O}(1-100\,\textrm{TeV})$. We can also take the dimensionless coefficients $\kappa_{11,12}$ and $\tilde{\kappa}_{22,12}$ small enough to suppress the VEVs $\langle\phi_{1,2}^{}\rangle$ in the case the Higgs bidoublet $\Phi$ is at an accessible scale $M_{\phi_{1,2}}^{}=\mathcal{O}(\textrm{TeV})$ while the PQ symmetry is still constrained to break at a high scale $\langle\sigma\rangle \gtrsim \mathcal{O}(10^{10}_{}\,\textrm{GeV})$.

The allowed Yukawa couplings are
\begin{eqnarray}
\!\!\!\!\!\!\mathcal{L}_Y^{}\!\!&=&\!\!-y_l^{}\bar{l}_{L}^{}\Phi l_{R}^{}\!-\!f(\bar{l}_L^{}\chi_{L2}^{}\xi_R^{}\!+\!\bar{l}_R^c\chi_{R2}^\ast \xi_R^{})
\!-\!y_E^{}(\bar{l}_L^{}\tilde{\chi}_{L2}^{}E_R^{}\nonumber\\
[2mm]
\!\!\!\!\!\!\!\!&&+\!\bar{l}_R^c\tilde{\chi}_{R2}^\ast E_L^{c})\!-\!f_E^{}\sigma\bar{E}_L^{}E_R^{}\!-\!y_q^{}\bar{q}_{L}^{}\Phi q_{R}^{}\!-\!y_U^{}(\bar{q}_L^{}\chi_{L1}^{}U_R^{}\nonumber\\
[2mm]
\!\!\!\!\!\!\!\!&&+\!\bar{q}_R^c\chi_{R1}^\ast U_L^{c})\!-\!f_U^{}\sigma\bar{U}_L^{}U_R^{}\!-\!y_D^{}(\bar{q}_L^{}\tilde{\chi}_{L2}^{}D_R^{}\nonumber\\
[2mm]
\!\!\!\!\!\!\!\!&&+\!\bar{q}_R^c\tilde{\chi}_{R2}^\ast D_L^{c})\!-\!f_D^{}\sigma\bar{D}_L^{}D_R^{}\!+\!\textrm{H.c.}\nonumber\\
[2mm]
\!\!\!\!\!\!\!\!&&\textrm{with}~~f_{U(D,E)}^{}=f_{U(D,E)}^{T},~y_{q(l)}^{}=y_{q(l)}^{T}.
\end{eqnarray}
The fermion masses thus should be
\begin{eqnarray}
\mathcal{L}\!\!&\supset&\!\!-\frac{1}{2}[\!\begin{array}{lll}\bar{\nu}_L^{}\!&\bar{\nu}_R^{c}\!&\bar{\xi}_R^{c} \end{array}\!]\!\!\left[\!\begin{array}{ccc}0&y_l^{}\langle\phi_1^{}\rangle& f \langle \chi_{L2}^{}\rangle\\
[2mm]
y_l^T\langle\phi_1^{}\rangle & 0 & f\langle \chi_{R2}^{}\rangle\\
[2mm]
f_{}^T\langle \chi_{L2}^{}\rangle & f_{}^T\langle \chi_{R2}^{}\rangle & 0 \end{array}\!\right]\!\!\!\!\left[\!\begin{array}{lll}\nu_L^{c}\\
[2mm]
\nu_R^{}\\
[2mm]
\xi_R^{} \end{array}\!\right]\nonumber\\
[2mm]
\!\!&&\!\!-[\!\begin{array}{ll}\bar{e}_L^{}\!&\bar{E}_L^{} \end{array}\!]\!\!\left[\!\begin{array}{ccc}y_l^{}\langle\tilde{\phi}_2^{}\rangle& y_E^{} \langle \tilde{\chi}_{L2}^{}\rangle\\
[2mm]
y_E^{T} \langle \tilde{\chi}_{R2}^{\dagger}\rangle &f_E^{}\langle \sigma\rangle\end{array}\!\right]\!\!\!\left[\!\begin{array}{ll}e_R^{}\\
[2mm]
E_R^{} \end{array}\!\right]\nonumber\\
[2mm]
\!\!&&\!\!-[\!\begin{array}{ll}\bar{u}_L^{}\!&\bar{U}_L^{} \end{array}\!]\!\!\left[\!\begin{array}{ccc}y_q^{}\langle\phi_1^{}\rangle& y_U^{} \langle \chi_{L1}^{}\rangle\\
[2mm]
y_U^{T} \langle \chi_{R1}^{\dagger}\rangle &f_U^{}\langle \sigma\rangle\end{array}\!\right]\!\!\!\left[\!\begin{array}{ll}u_R^{}\\
[2mm]
U_R^{} \end{array}\!\right]\nonumber\\
[2mm]
\!\!&&\!\!-[\!\begin{array}{ll}\bar{d}_L^{}\!&\bar{D}_L^{} \end{array}\!]\!\!\left[\!\begin{array}{ccc}y_q^{}\langle\tilde{\phi}_2^{}\rangle& y_D^{} \langle \tilde{\chi}_{L2}^{}\rangle\\
[2mm]
y_D^{T} \langle \tilde{\chi}_{R2}^{\dagger}\rangle &f_D^{}\langle \sigma\rangle\end{array}\!\right]\!\!\!\left[\!\begin{array}{ll}d_R^{}\\
[2mm]
D_R^{} \end{array}\!\right]+\textrm{H.c.}\,.
\end{eqnarray}
By block diagonalizing the above mass matrices, we have
\begin{eqnarray}
\!\!\!\!\mathcal{L}\!\!&\supset&\!\!-M_{N}^{}\bar{\nu}_R^{c}\xi_R^{}-\frac{1}{2}m_\nu^{}\bar{\nu}_L^{}\nu_L^c~~\textrm{with}\nonumber\\
[2mm]
\!\!\!\!\!\!&&\!\!M_N^{}=f\langle\chi_{R2}^{}\rangle\,,~~m_\nu^{}=-2y_l^{}\langle\phi_1^{}\rangle\frac{\langle\chi_{L2}^{}\rangle}{\langle\chi_{R2}^{}\rangle},
\end{eqnarray}
for the neutral fermions, and
\begin{eqnarray}
\!\!\!\!\mathcal{L}\!\!&\supset&\!\!-m_u^{}\bar{u}_L^{}u_R^{}-m_d^{}\bar{d}_L^{}d_R^{}-m_e^{}\bar{e}_L^{}e_R^{}-M_U^{}\bar{U}_L^{}U_R^{}\nonumber\\
[2mm]
\!\!\!\!\!\!&&\!\!-M_D^{}\bar{D}_L^{}D_R^{}-M_E^{}\bar{E}_L^{}E_R^{}+\textrm{H.c.}~~\textrm{with}\nonumber\\
[2mm]
\!\!\!\!M_{U}^{}\!\!&=&\!\!f_{U}^{}\langle\sigma\rangle\,,
~~m_u^{}=y_q^{}\langle\phi_1^{}\rangle-y_U^{}\frac{\langle\chi_{L1}^{}\rangle\langle\chi_{R1}^\dagger\rangle}{M_U^{}}y^T_U\,;\nonumber\\
[2mm]
\!\!\!\!M_{D}^{}\!\!&=&\!\!f_{D}^{}\langle\sigma\rangle\,,
~~m_d^{}=y_q^{}\langle\tilde{\phi}_2^{}\rangle-y_D^{}\frac{\langle\tilde{\chi}_{L2}^{}\rangle\langle\tilde{\chi}_{R2}^\dagger\rangle}{M_D^{}}y^T_D\,;\nonumber\\
[2mm]
\!\!\!\!M_{E}^{}\!\!&=&\!\!f_{E}^{}\langle\sigma\rangle\,,
~~m_e^{}=y_l^{}\langle\tilde{\phi}_2^{}\rangle-y_E^{}\frac{\langle\tilde{\chi}_{L2}^{}\rangle\langle\tilde{\chi}_{R2}^\dagger\rangle}{M_E^{}}y^T_E\,,
\end{eqnarray}
for the charged fermions. Compared with the fermion masses in the revived original LRSM, the neutrino masses are still induced by the linear seesaw, while the SM charged fermion masses additionally have a universal seesaw \cite{balakrishna1988,bkm1988} contribution (the second terms in the charged fermion masses $m_{u,d,e}^{}$). It is easy to find that for $\langle\phi_{1,2}\rangle=\mathcal{O}(1-100\,\textrm{eV})$, $\langle\chi_{L1,2}^{}\rangle<\langle\varphi\rangle=174\,\textrm{GeV}$ and $\langle\chi_{R1,2}\rangle=\mathcal{O}(1-100\,\textrm{TeV})$, the neutrino masses can naturally arrive at the sub-eV scale while the charged fermion masses should mostly come from the universal seesaw.

The effective operators for the $0\nu\beta\beta$ processes are
\begin{eqnarray}
\!\!\!\!\!\!\!\!\mathcal{L}\!\!&\supset& \!\!16 G_F^2\bar{u}_L^{}\gamma^\mu_{}d_L^{}\bar{e}_L^{}\gamma_\mu^{}\frac{(m_\nu^{})_{ee}^{}}{q^2_{}}\gamma_\nu^{}e_L^c \bar{u}_L^{}\gamma^\nu_{}d_L^{} \nonumber\\
[2mm]
\!\!\!\!\!\!\!\!\!\!&&\!\!+8\frac{\langle\chi_{L2}^{}\rangle}{\langle\chi_{R2}^{}\rangle}\frac{M_{W_L^{}}^{2}}{M_{W_R^{}}^{2}}
G_F^2\bar{u}_L^{}\gamma^\mu_{}d_L^{}\bar{e}_L^{}\gamma_\mu^{}\frac{q\!\!\!\diagup}{q^2_{}}\gamma_\nu^{}e_R^c \bar{u}_R^{}\gamma^\nu_{}d_R^{}\,.
\end{eqnarray}
The half-life of the $0\nu\beta\beta$ processes can be computed by replacing the factor $\langle\chi_L^{}\rangle^2_{}/\langle\chi_{R}^{}\rangle^2_{}$ to $\langle\chi_{L2}^{}\rangle^2_{}/\langle\chi_{R2}^{}\rangle^2_{}$ in Eq. (\ref{prediction}). As an example, we can take the seesaw condition $\langle\chi_{R2}^{}\rangle\gg \langle\chi_{L2}^{}\rangle$ to be $\langle\chi_{R2}^{}\rangle^2_{}=1000\,\langle\chi_{L2}^{}\rangle^2_{}$, and then perform $T_{1/2}^{0\nu}(_{}^{136}\textrm{Xe})>8\times 10^{26}_{}\,\textrm{yr}$ for $M_{W_R^{}}^{}> 32.4-36.9\,\textrm{TeV}$ and $T_{1/2}^{0\nu}(_{}^{76}\textrm{Ge})>6\times 10^{27}_{}\,\textrm{yr}$ for $M_{W_R^{}}^{}> 32.5-47.6\,\textrm{TeV}$. With this parameter choice, the mixing between the left-handed neutrinos and the singlet fermions will not conflict with the precise measurement of the invisible decay width of the $Z$ boson even if the Dirac pairs composed of the right-handed neutrinos and the singlet fermions are heavier then the $Z$ boson. Alternatively, we can consider a little weaker seesaw condition $\langle\chi_{R2}^{}\rangle=10\,\langle\chi_{L2}^{}\rangle$ by assuming the right-handed neutrinos and the singlet fermions much lighter than the $Z$ boson. In this case, we can have $T_{1/2}^{0\nu}(_{}^{136}\textrm{Xe})<8\times 10^{26}_{}\,\textrm{yr}$ for $M_{W_R^{}}^{}<57.7-65.8\,\textrm{TeV}$ and $T_{1/2}^{0\nu}(_{}^{76}\textrm{Ge})<6\times 10^{27}_{}\,\textrm{yr}$ for $M_{W_R^{}}^{}<57.8-84.7\,\textrm{TeV}$.

The present model has some advantages. Firstly, it can automatically forbid the Majorana mass term of the singlet fermions and naturally suppress the Dirac mass term between the left- and right-handed neutrinos. We hence need not artificially take the essential assumptions in the previous demonstrations. Secondly, the light quarks $u,d$ and the heavy quarks $U,D$ have the axial couplings to the Goldstone boson $a$ as follows,
\begin{eqnarray}
\!\!\!\!\mathcal{L}\supset\frac{\partial_\mu^{}a}{f_{PQ}^{}}\left(\bar{u}\gamma^\mu_{}\gamma_5^{}u+\bar{d}\gamma^\mu_{}\gamma_5^{}d
+
\bar{U}\gamma^\mu_{}\gamma_5^{}U+\bar{D}\gamma^\mu_{}\gamma_5^{}D\right).
\end{eqnarray}
The Goldstone boson $a$ will contribute to the strong CP phase through the DFSZ \cite{zhitnitsky1980,dfs1981} scheme (the axial couplings to the light quarks $u,d$) and the KSVZ \cite{kim1979,svz1980} scheme (the axial couplings to the heavy quarks $U,D$). Therefore, the Goldstone boson $a$ is an invisible axion and the global symmetry $U(1)_{PQ}^{}$ is a PQ symmetry. The PQ symmetry breaking scale should have a lower limit $f_{PQ}^{}\gtrsim 10^{10}_{}\,\textrm{GeV}$ to avoid the astrophysical constraints. The axion $a$, which eventually picks up a tiny mass through the color anomaly and then becomes a pseudo Goldstone, can act as a cold dark matter particle for a proper choice of the breaking scale of the PQ symmetry and the initial value of the strong CP phase \cite{olive2014}. Thirdly, the $[SU(2)\times SU(2)_R^{}]$-bidoublet Higgs scalar $\Phi$ has a seesaw-suppressed VEV. So its Yukawa couplings to the quarks can only give a negligible contribution to the quark masses and hence their values can be set flexibly. Meanwhile, the Yukawa couplings of this Higgs bidoublet to the leptons are completely determined by the neutrino mass matrix. The Higgs bidoublet is allowed near the TeV scale, so that it may be tested at the running and/or planning colliders. In particular the decays of the Higgs bidoublet into the charged leptons can open a window to measure the neutrino mass matrix.

\section{Summary}

In this paper we have shown in the LRSMs for the linear seesaw, a $0\nu\beta\beta$ process induced by the left- and right-handed charged currents together decouples from any masses and mixing of the left- and right-handed neutrinos, but depends on the VEVs of the Higgs scalars for driving the left-right and electroweak symmetry breaking. This $0\nu\beta\beta$ process can reach the experimental sensitivities if the right-handed charged gauge boson is below the 100 TeV scale. In our realistic model, the linear seesaw can be verified at colliders if the related Higgs bidoublet is close to the TeV scale.

\textbf{Acknowledgement}: This work was supported by the Shanghai Jiao Tong University under Grant No. WF220407201, the Recruitment Program for Young Professionals under Grant No. 15Z127060004 and the Shanghai Laboratory for Particle Physics and Cosmology under Grant No. 11DZ2260700.

\end{document}